\documentclass[twocolumn,showpacs,preprintnumbers,amsmath,amssymb]{revtex4}
\usepackage{graphicx,epsfig}
\usepackage{dcolumn}
\usepackage{bm}

\begin{document}

\title{ Contact Pair Dynamics During Folding of a Model Globular Protein, Hp-36}
 
\author{Arnab Mukherjee and Biman Bagchi}
\affiliation{Solid State and Structural Chemistry Unit,
Indian Institute of Science,
Bangalore, India 560 012.}
\email{bbagchi@sscu.iisc.ernet.in}

\date{\today}

\begin{abstract}
The dynamics of contact pair formation between various hydrophobic residues during 
folding of a model protein Hp-36 is investigated by Brownian dynamics simulation. 
Hydropathy scale and non-local helix propensity of amino acids are used to model the 
complex interaction potential. The resulting structure of the model protein mimics 
the native state of the real protein with a $RMSD$ of 4.5 \AA. A contact pair distance 
time correlation function (CPCF), $C_{P}^{ij}(t)$, is introduced which shows multistage 
decay, including a {\it slow late stage dynamics} for a few specific pairs. 
{\it These pairs determine the long time folding rate}. Dynamics can be correlated 
with the landscape, relative contact order and topological contact.
\end{abstract}


\maketitle
%****************************
%\section {Introduction}
%****************************
 The dynamics of folding of an extended protein chain at high temperature (or high 
urea concentration) to its unique folded state at low temperature (or low urea 
concentration) is a highly complex problem with many interesting aspects. 
Recent experimental, theoretical and computer simulations studies $[1-7]$ 
have unearthed and explained many fascinating aspects of folding, although still 
many others remain to be 
explored. The paradigm of landscape (with the idea of folding funnel) has 
provided new insight into the problem \cite{pgscience,bw}. 
Experimental data on the rate of folding of a large number of small proteins have 
suggested a close relation between the the relative contact order and the rate of 
folding \cite{baker}. The relative contact order denotes the average sequence 
distance between the hydrophobic pair contacts and is defined as \cite{plaxo},
\begin{equation}
RCO = {{\sum_{i,j} (s_{j} - s_{i})}\over{L N_{c}}}
\end{equation}
\noindent where ($i,j)$ are the specific hydrophobic pair contacts, $N_{c}$ is the 
number of contacts while $L$ is total number of hydrophobic amino acids present in 
the protein. $s_{j}$ and $s_{i}$ are the sequence number along the contour of the 
chain. The rate of folding was found to decrease nearly exponentially with $RCO$. 
The dynamics of such non-local contact formation holds the key to the understanding 
of dynamics of folding. However, this aspect has remained largely unexplored. 

    An attractive way to explore pair dynamics of hydrophobic contact is via the 
technique of fluorescence resonance energy transfer (FRET). In FRET, one measures the 
time dependence of energy transfer from a chosen donor fluorophore to a chosen acceptor. 
The rate of transfer may be due to dipolar interactions and the rate of transfer is 
given by the well known F\"{o}rster expression \cite{forster},
\begin{equation}
k_{f} = k_{rad} \left( {{R_{F}}\over {R}}\right)
\end{equation}
\noindent where $k_{rad}$ is the radiative rate and $R_{F}$ is the  F\"{o}rster radius. By 
suitably choosing donor-acceptor pair, $R_{F}$ can be varied over a wide range. This 
allows the study of the dynamics of pair separation, essential to understand protein 
folding \cite{gray}. $k_{rad}$ is typically less than 
(but of the order of) $10^{9}$ $sec^{-1}$. Thus F\"{o}rster transfer provides us with 
a sufficiently fast camera to take snapshots of the dynamics of contact pair formation.

 In this study, we have studied contact formation by Brownian dynamics simulations of a 
model protein Hp-36 which is one of the smallest protein that 
folds autonomously to a stable compact ordered structure, with a large helix content 
\cite{mcknight}. Hp-36 is a subdomain of chicken villin which is implicated in the 
formation of microvilli in the absorbtive epithelium of the gut and the proximal tube 
of the kidney \cite{bretscher}. All atom simulation study on this Hp-36 have revealed 
at least two pathways of folding \cite{kollman}. Earlier, several studies of Hp-36 
were presented using Monte Carlo technique \cite{hansmannprl} and Brownian dynamics 
\cite{srini}.
%*************************************
%\section{Construction and Energy Functions of the model proteins}
%*************************************

 The model studied here is constructed by taking two atoms 
for a particular amino acid. The smaller atom represents the  backbone 
$C_\alpha$ atom of real protein while the bigger atom mimics the whole side 
chain residue. Construction of the model protein has been described 
in detail elsewhere\cite{arnjcp}. Similar types of model (with more rigorous force 
field) have been introduced by Scheraga et al. recently \cite{scheraga}. 
The total potential energy function of the model protein $V_{Total}$ is written as,
\begin{equation}
V_{Total} = V_B + V_\theta + V_T + V_{LJ} + V_{helix}
\end{equation}
\noindent where $V_B$ and $V_\theta$ are the potential contributions due to vibration 
of bonds and bending motions of the bond angles. Standard harmonic potential is assumed 
for the above two potentials with spring constants 43.0 kJ mole$^{-1}$\AA$^{-2}$ and 
8.6 kJ mole$^{-1}$\AA$^{-2}$ for the bonds between backbone atoms and bonds 
joining side residues with the backbone atoms, respectively. In case of the 
bending potential, spring constant is taken to be  10.0 kJ mol$^{-1}$ rad$^{-2}$. 
$V_{T}$ ($ = \epsilon_{T} \sum_{\phi} (1/2) [1 + cos(3\phi)]$) is taken as torsional 
potentials for the rotations of the bonds. $\epsilon_{T}=$ 1 kJ mol$^{-1}$.
\begin{figure}[htb]
\epsfig{file=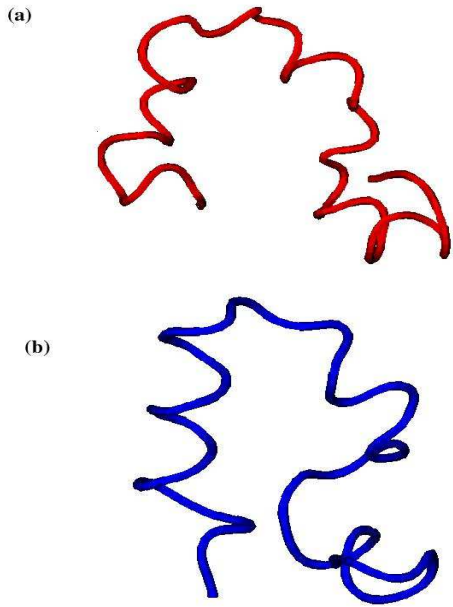,height=8cm,width=6cm}
\caption{\label{fig1}(a) The backbone structure of the model protein with 
the lowest $RMSD$ (4.5 \AA). (b) The backbone structure of the 
native state of real Hp-36.}
\end{figure}
 The nonbonding potential $V_{LJ}$ is the sum of the pair interactions between the 
atoms and is given by,
\begin{eqnarray}
V_{LJ} = 4\sum_{i,j}\epsilon_{ij}\biggl [\biggl({\sigma_{ij}\over r_{ij}}
\biggr )^{12} - \biggl({\sigma_{ij}\over r_{ij}}\biggr)^6 \biggr ]
\end{eqnarray}
\noindent where $r_{ij}$ and $\epsilon_{ij}$ are the distance and interaction between the 
$i$-th and $j$-th atom. $\sigma_{ij} = {1\over 2}(\sigma_{ii} + \sigma_{jj})$ and 
$\epsilon_{ij} = \sqrt{\epsilon_{ii}\epsilon_{jj}}$. Sizes and interactions are taken 
to be the same (1.8 \AA \, and 0.05 kJ mol$^{-1}$, respectively) for all the backbone 
atoms as they represent the $C_{\alpha}$ atoms in case of real proteins. Side residues, 
on the other hand, carry the characteristics of a particular amino acid. Different sizes 
of the side residues are taken from the values given by Levitt \cite{levwar}. 
Interactions of the side residues are obtained from the hydrophobicities of the amino 
acids. We construct effective potential guided by the well-known statistical mechanical relation 
between potential of mean force and radial distribution function as 
$V_{ij}^{eff} = -k_{B}T ln g_{ij}(r)$  \cite{hansen}. Strong 
correlation among the hydrophobic groups (absent among the hydrophilic amino acids) 
implies that the hydrophobic amino acids should have 
stronger effective interaction than the hydrophilic groups. So the interaction 
parameters of the side residues can be mapped from the hydropathy scale \cite{kyte} by 
using a linear equation as given below,
\begin{equation}
\epsilon_{ii} = \epsilon_{min} + (\epsilon_{max} - \epsilon_{min}) * 
\biggl({H_{ii} - H_{min}\over{H_{max} - H_{min}}}\biggr)
\end{equation}
\noindent where, $\epsilon_{ii}$ is the interaction parameter of the 
$i$th amino acid with itself. $\epsilon_{min}$(=0.2 kJ mol$^{-1}$) and 
$\epsilon_{max}$(=11.0 kJ mol$^{-1}$) are the 
minimum and maximum value of the interaction strength chosen for the most 
hydrophilic(arginine) and most hydrophobic(isoleucine) amino acids, respectively. 
$H_{ii}$ is the hydropathy index of $i$th amino acid given by Kyte and 
Doolittle \cite{kyte} and $H_{min}$(=-4.5) and $H_{max}$(=4.5) are the minimum 
and maximum hydropathy index among all the amino acids. Further details 
are available in Ref. $16$.
\begin{figure}[htb]
\epsfig{file=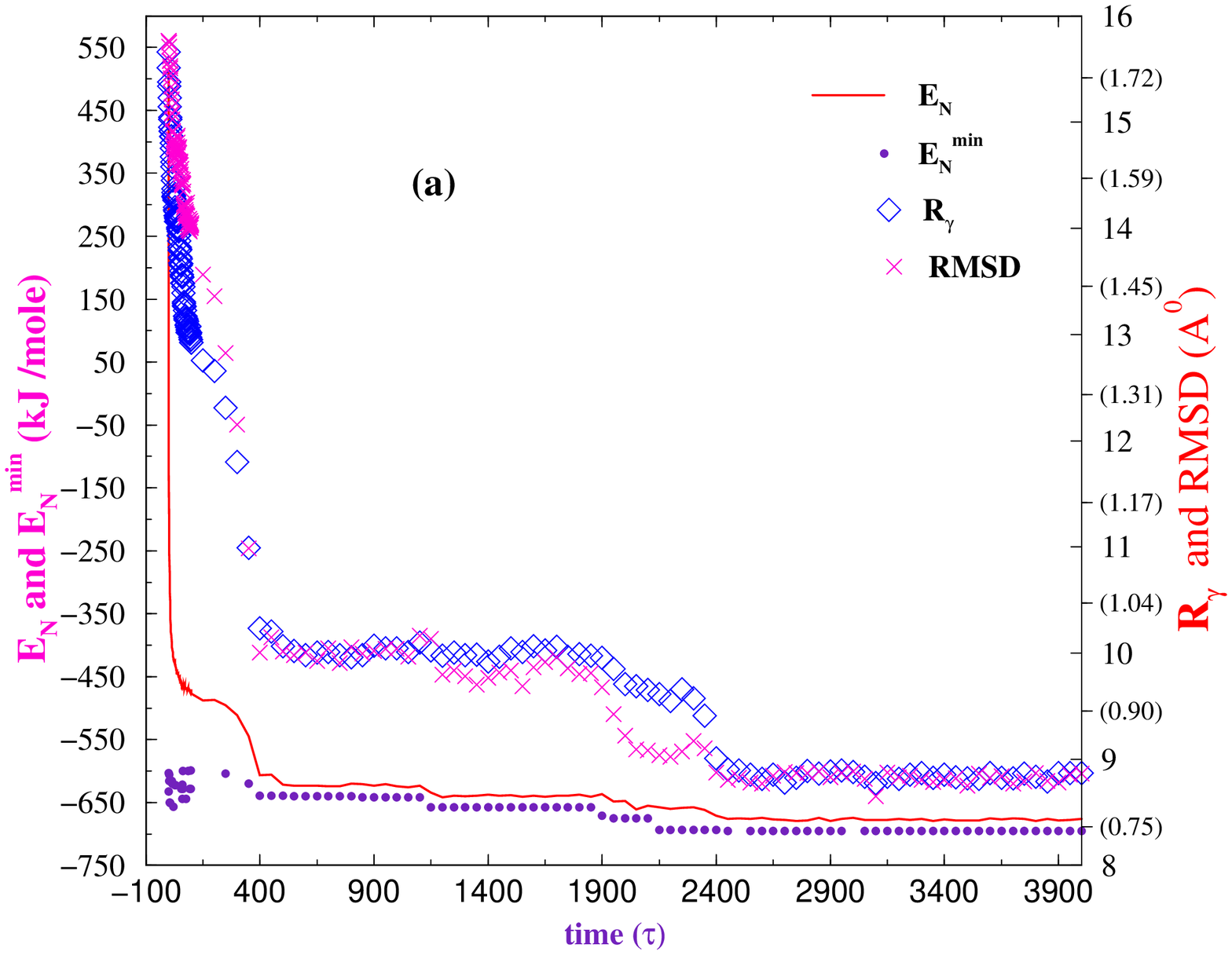,height=4cm,width=6cm}
\epsfig{file=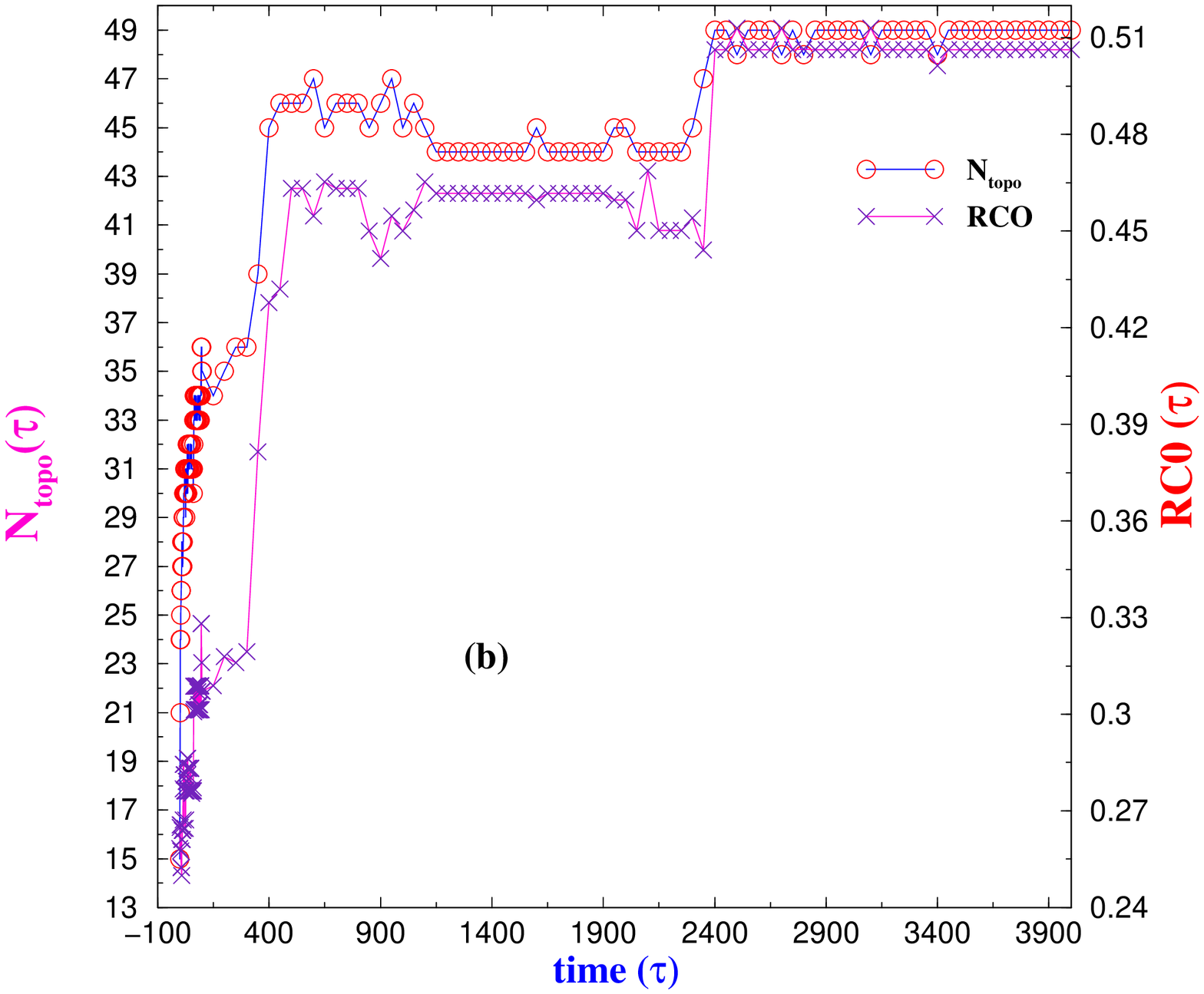,height=4cm,width=6cm}
\caption{\label{fig2}(a) The multistage temporal evolution of energy $E_{N}$ and minimized 
energy $E_{N}^{min}$ (left ordinate), radius of gyration $R_{\gamma}$ 
(right ordinate) and $RMSD$ (right ordinate in parenthesis) are plotted in the 
same time axis to show the dynamical correlation. (b) Dynamics of the topological 
contact $N_{topo}$ (left ordinate) and the $RCO$ (right ordinate) are plotted in the 
same time axis which show a similar dynamical growth.}
\end{figure}
  An important part of secondary structure of the real protein is the formation of 
$\alpha$ helix. In the absence of hydrogen bonding, we introduce the following 
effective potential among the backbone atoms to mimic the helix formation 
along the chain of residues that show high helix propensity,
\begin{equation}
V_{helix} = \sum_{i=3}^{N-3}\biggl [{1\over 2} K_{i}^{1-3} (r_{i,i+2} - r_{h})
^2 + {1\over 2} K_{i}^{1-4} (r_{i,i+3} - r_{h})^2 \biggr ]
\end{equation}
where $r_{i,i+2}$ and $r_{i,i+2}$ are the distances of $i$th atom with $i+2$ and 
$i+3$ th atoms, respectively. $r_{h}$ is the equilibrium distance and is taken 
as 5.5 \AA, motivated by the observation that the distance of $r_{i}$ with 
$r_{i+2}$ and $r_{i+3}$ is nearly constant at 5.5 \AA \, in an $\alpha$ helix. 
The summation excludes the first and last three amino acids as there is less 
helix formation observed in the ends of the protein chain \cite{zimm}.
The force constant for the above harmonic potential is mapped from the helix 
propensities $Hp_{i}$ taken from Scholtz {\it et al.} \cite{pace}, 
${\cal K}_{i} = {\cal K}_{alanine} - Hp_{i} \times ({\cal K}_{alanine} - 
{\cal K}_{glycine})$. ${\cal K}_{alanine}$ and ${\cal K}_{glycine}$ 
are the force constants for alanine and glycine, 17.2 and 0.0 kJ mol$^{-1}$, 
respectively. Next, the influence of the neighboring amino acids for the formation of 
helix has been considered by taking an average of the spring constants as 
$ K_{i}^{1-3} = {1\over 3} [{\cal K}_{i} + {\cal K}_{i+1} + {\cal K}_{i+2}] $ and 
$K_{i}^{1-4} = {1\over 4} [{\cal K}_{i} + {\cal K}_{i+1} + {\cal K}_{i+2} + 
{\cal K}_{i+3}]$, with the condition that $K_{i}^{1-3},K_{i}^{1-4}\ge 0$ as the force 
constant must remain positive. The above formulation of helix potential is motivated 
by the work of Chou and Fasman about the prediction of helix formation that 
{\it the neighbors of a particular amino acid} should be considered rather than 
its own helix propensity \cite{chou}. 

 Figure $1(a)$ shows the best folded structure obtained in our simulations while figure 
$1(b)$ shows the real one. The $RMSD$ calculated over backbone with the real native 
Hp-36 is 4.5 \AA \, which is reasonable for such a simplified model.
%*********************************************
%\section{Simulation Detail}
%*********************************************
 The initial configuration of the model protein was generated by configurational 
bias Monte Carlo technique \cite{frenkel}. Atoms attached to a single branch point 
were generated simultaneously. Then the initial configuration was subjected to 
Brownian dynamics simulation for the study the folding. Time evolution of the model 
protein was carried out according to the motion of each bead as below,
\begin{equation}
{\bf r}_{i}(t+\Delta t) = {\bf r}_{i}(t) + {D_{i}\over {k_{B}T}}
{\bf F}_{i}(t)\Delta t + \Delta {\bf r}_{i}^{G}
\end{equation}
\noindent where each component of $\Delta {\bf r}_{i\alpha}^{G}$ is taken from 
a Gaussian distribution with mean zero and variance $\langle 
(\Delta r_{i \alpha}^{G})^{2}\rangle = 2 D \Delta t$ \cite{mccammon,hansen}. 
${\bf r}_{i}(t)$ is the position of the $i$th atom at time t and the systematic 
force on $i$th atom at time t is ${\bf F}_{i}(t)$.  The time step $\Delta t$ is taken 
as 0.001. $D_{i}$ is the diffusion coefficient of the $i$-th particle calculated from 
the Stokes-Einstein relation $D_{i} = {{k_{B}T}\over{6\pi\eta R_{i}}}$. 
$R_{i}$ is the radius of the $i$-th atom and $\eta$ is the viscosity of 
the solvent. $k_{B}$ and $T$ are the Boltzmann constant and temperature, 
respectively. Simulations have been carried out for ${\cal N}$ number of different 
initial configurations, where ${\cal N} = 584$.   
%***************************************************
%\section{Results and Discussion}
%***************************************************

 The potential energy $E_{N}$, the radius of gyration $R_{\gamma}$ and the $RMSD$ 
(calculated over backbone from the real native structure of Hp-36 obtained from protein 
data bank \cite{pdb,mcknight}) all exhibit the multistage dynamics as shown in figure 
$2(a)$ in the same time axis. 
There is an initial sharp hydrophobic collapse followed by a slower 
decay is observed till 500$\tau$ ($\approx$ 200 ns). A long plateau follows in the final 
stage which exists for a very long time ($~$2000$\tau \approx$ 1$\mu$s) before the 
protein reaches its final lowest energy state at around 2400$\tau$. Energy values of the 
corresponding inherent structures obtained by conjugate gradient is shown by the 
symbols {\it which depicts the decreasing local minima attained by the system until it 
reaches the final folded state}. Total number of hydrophobic topological contact 
$N_{topo}$ and relative contact order $RCO$ are plotted in figure $2(b)$. $N_{topo}$ 
is defined to be formed if two hydrophobic side chain residues come within a distance 
of 8.5 \AA. $RCO$ is calculated from Eq. $1$. Both $N_{topo}$ and $RCO$ show a multistage 
increase signifying the participation of nonlocal contacts.
\begin{figure}[htb]
\epsfig{file=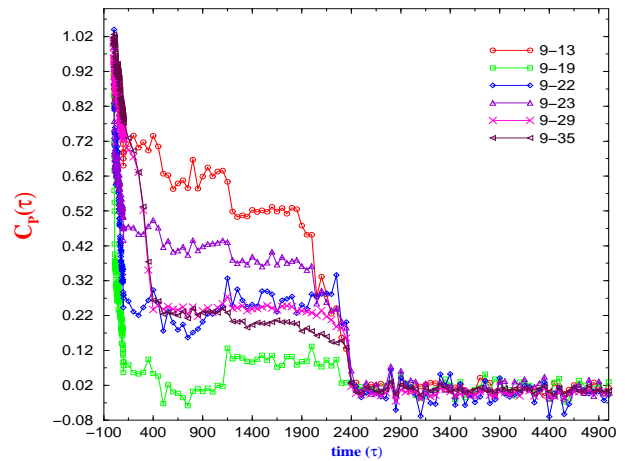,height=6cm,width=8cm}
\caption{\label{fig3} Dynamics of contact formation of different side residues 
with the $9$-th side residue is shown. The multistage relaxation process in the 
dynamical quantities originates from the diverse dynamics of the contact pairs.}
\end{figure}

 Folding can be probed microscopically by monitoring the dynamics of 
separation between different amino acid pairs. The widely different time scales of 
movement of all the different pairs together give rise to an overall picture of the 
dynamics of folding which is reflected in the macroscopic quantities. The effective 
dynamics of pair separation can be described by introducing a new pair correlation 
function defined below \cite{srini},
\begin{equation}
C_{P}^{ij}(t) = {{d^{ij}(t) - d^{ij}(\infty)}\over{d^{ij}(0)-d^{ij}(\infty)}}
\end{equation}
\noindent where, $d^{ij}(t) = |{\bf r}_{i}(t) - {\bf r}_{j}(t)|$. 
${\bf r}_{i}$ and ${\bf r}_{j}$ are the 
positions of the $i$-th and $j$-th atom, respectively. 
Figure $3$ shows the $C_{P}^{ij}(t)$ of the $9$th side residue with many 
other hydrophobic side residues. The three stages of the folding process are 
reflected by mainly three different dynamical behavior seen amongst the side residues. 
Side residues closed to the tagged one collapse very fast. Some show 
an initial shoulder and only a few show the plateau in the long time decay {\it that 
correlates with the similar plateau observed in case of other dynamical quantities}.
The final decrease in energy is observed when all the different contact pair correlation 
functions decay at around 2400$\tau$. 

 We have calculated survival probability $S_{P}(t)$ in FRET using F\"{o}rster energy 
transfer rate from Eq. $2$ for the $9-35$ pair along the Brownian dynamics 
trajectory leading to the most stable structure. The F\"{o}rster radius $R_{F}$ is 
taken as 10 \AA. Figure $4$ shows initial very slow decrease of $S_{P}(t)$ to be 
followed by a sudden drop at around 2400$\tau$ as seen in case of the different 
dynamical properties discussed above. $S_{P}(t)$ is found to be relatively insensitive 
to $k_{rad}$.
\begin{figure}[htb]
\epsfig{file=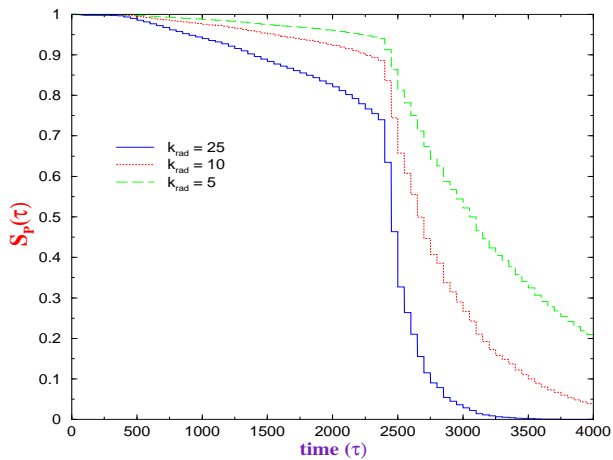,height=6cm,width=8cm}
\caption{\label{fig4} FRET survival probability for $R_{F}=10.0$ \AA \, is plotted 
for different radiative rates to capture the wide variety of time scales involved in FRET.}
\end{figure}
%**********************
%\section {Conclusion}
%**********************
The resemblance with the real native state of the protein and the dynamics of $RMSD$ 
show the validity of the model used for this protein. Contact pair dynamics and the 
time evolution of energy, radius of gyration, relative contact order formation etc. 
brings out the rich and diverse dynamics of protein folding. The initial ultrafast 
hydrophobic collapse signify that the upper part of the funnel is steep -- followed 
by a change in slope. {\it Rate determining step, however, arises from the final stage 
of folding on a very flat and rugged underlying landscape marked by the large 
conformational entropy barrier with little energy change} \cite{bob}. This entropic 
bottleneck arises from the necessity to form long range hydrophobic contacts, as 
envisaged by Dill and Wolynes. The atoms mimicking the whole side chain of the real 
protein play a very important role for structural and dynamical aspects in this study. 
Moreover, the new contact pair correlation function and FRET probes the folding events 
in minute detail. 
%{\bf Acknowledgement}

 A. Mukherjee thanks Ashwin, Kausik and Prasanth for technical discussions. 
Authors thank CSIR, New Delhi, India and DST, India for financial support.

%{\large \bf Figure Captions}
\end{document}